\def \manuflag {0}
 
\ifnum \manuflag = 0
 \documentclass[pre,twocolumn,showpacs,amsmath,amssymb]{revtex4}
 \usepackage{psfig}	
 \usepackage{rotating}	
 \usepackage{epsfig}		
 \usepackage{delarray}
 \usepackage{graphicx}
 \usepackage{dcolumn}
 \usepackage{bm}
 \def \Title{
Large-scale Oscillation of
Structure-Related DNA Sequence Features in Human Chromosome 21
}

 \newcommand{\SEC}{\section}
 
\else
 \documentclass[pre,onecolumn,showpacs,amsmath,amssymb]{revtex4}
 \usepackage{psfig}	
 \usepackage{epsfig}		
 \usepackage{delarray}
 \usepackage{graphicx}
 \usepackage{dcolumn}
 \usepackage{bm}
 \usepackage{rotating}
 \def \Title{
Large-scale Oscillation of
Structure-Related DNA Sequence Features in Human Chromosome 21 }

\fi

\def \Abstract{
Human chromosome 21 is the only chromosome in human
genome that exhibits oscillation of (G+C)-content
of cycle length of hundreds kilobases (500 kb near
the right telomere). 
We aim at establishing the existence of similar periodicity in
structure-related sequence features in order to 
relate this (G+C)\% oscillation to other biological phenomena.
The following quantities are shown to oscillate with 
the same 500kb periodicity in human chromosome 21: 
binding energy calculated by two sets of dinucleotide-based 
thermodynamic parameters,
AA/TT and AAA/TTT bi-/tri-nucleotide density,
5'-TA-3' dinucleotide density, and signal for 10/11-base 
periodicity of AA/TT or AAA/TTT.
These intrinsic  quantities are related to structural
features of the double helix of DNA molecules, such
as base-pair binding, untwisting/unwinding,  stiffness,
and a putative tendency for nucleosome formation.
}

\ifnum \manuflag = 0
 
\begin{document}
 \title{\Title}
 \author{Wentian Li}
 \email{wli@nslij-genetics.org}
 \affiliation{The Robert S. Boas Center for Genomics and Human Genetics,
Feinstein Institute for Medical Research, North Shore LIJ Health System,
350 Community Drive, Manhasset, NY, USA. 
	      	}
 \author{Pedro Miramontes}
 \email{pmv@fciencias.unam.mx}
 \affiliation{
Departamento de Matem\'{a}ticas,
Facultad de Ciencias, Universidad Nacional Aut\'{o}noma de M\'{e}xico,
Circuito Exterior, Ciudad Universitaria, 04510 M\'{e}xico, D.F. and \\
Departamento de Matem\'{a}ticas, Universidad de Sonora,
Encinas y Rosales, Hermosillo 83000 Sonora, M\'{e}xico.
 }
 \begin{abstract}
   \Abstract
 \end{abstract}
 %% \keywords{Suggested keywords if desired}
 \maketitle
\else
 \begin{document}
 \title{\Title}
 \author{Wentian Li}
 \email{...@...}
 \affiliation{...}
 \author{Pedro Miramontes}
 \email{pmv@fciencias.unam.mx}
 \affiliation{,
 }
 \begin{abstract}
   \Abstract
 \end{abstract}
 \pacs{87.10.+e, 02.50.-r, 05.40.-a  \hfill {\tt Thu Sep  5 15:36:39 EDT 2002}}
 %% \keywords{Suggested keywords if desired}
\maketitle
\fi

\SEC{Introduction}

DNA sequences are full of features at small, intermediate,
and large scales \citep{li97}. At short distances, there is strong
periodicity-of-three-nucleotide signal in protein-coding
regions (but absent in non-coding regions) \citep{period3},
and  a weaker but ubiquitous 10-11 bases signal in many genomes
\citep{period10}.  At intermediate length scale, there are 
{\sl Alu} sequences of about 300 bases long \citep{alu}, 
and nucleosome-forming sequence of around 120-200 bases 
\citep{nucleosome}. At large length scales, the most well known 
features are the existence of alternating (G+C)\%-high and
(G+C)\%-low ``isochores" \citep{isochore}, and the distribution of sine
wave that prefers long-wavelength signals (the so-called
``1/f" spectra when viewed in the spectral space
\citep{1f}).

A recent survey of (G+C)\% fluctuation in all human 
({\sl Homo sapiens}) chromosomes
revealed that chromosome 21 exhibits a unique
500 kilobases (kb) oscillation in (G+C)\% \citep{li-holste04}.
This oscillation starts around the position of 43.5 million
bases (Mb) and lasts five cycles (with five (G+C)\%-low
six (G+C)\%-high peaks). No other human chromosomes exhibit
similar periodicities with such a long cycle length.

Human chromosome 21 has other special properties as
compared to the rest of the human chromosomes. First, it is
the shortest human chromosome. Second, its (G+C)\% increases
stepwise from left (centromeric) to right (telomeric, i.e.,
close to the end of the chromosome),
with three distinct ``super" isochore regions (see, e.g. Fig.3
of \citep{isochore}(b)). The 500kb oscillation of
(G+C)\% described above appears in the third region
with the highest (G+C)\% and the highest gene content.
Third, the failure rate in segregating  homologous chromosomes
during meiosis is the highest among surviving infants 
in human chromosome 21 than any other human chromosomes. 
When this happens, the surviving infants typically carries 
three copies of chromosome 21 (``trisomy 21") instead 
of one copy \citep{trisomy}.
The resulting Down syndrome is the leading case
of birth defects \citep{patterson}.

The uniqueness of the 500kb oscillation in (G+C)\% in
human chromosome 21 and highest trisomy rate in chromosome
21 among surviving infants motivated us to speculate the
possibility that this 500kb oscillation might be somewhat
related to the trisomy risk. An argument is that
the periodicity in (G+C)\% is a basis for certain structural
periodicity, which in turn might interfere with the
proper segregation of chromatids during meiosis.
One intriguing observation is that for younger mothers
with trisomy 21, the placement of meiosis exchange
tends to be telomeric \citep{sherman05}.

In this paper, we examine whether sequence-based
structure features oscillates with the 500kb
cycle length in the telomeric region of human chromosome 21.
The structural features we focus on include
helix binding energy, flexibility or
stiffness in secondary structure of DNA helix,
tendency for nucleosome formation based on periodicity
of 10-11 bases, and a tendency for anchoring DNA loops.

Note that only the intrinsic quantities are
calculable here: chromatin structures that depend on
extrinsic protein factors require experimental
data, and these evidences are not yet conclusive. Also
note that the sequence-to-structure connections in
some model are based on simplified assumptions, and
our calculation may only give a partial picture of
DNA helix structure properties. Our hope is for this 
work to contribute to the eventual establishment
of a sequence-function connection.

\begin{table}[t]
\begin{center}
\begin{tabular}{lcccc}
5' /\ 3'  & G & A & T & C \\
\hline
G & 2.75/1.84 & 1.41/1.30 & 1.13/1.44 & 2.82/2.24 \\
A &  (see CT) & 1.66/1.00 & 1.19/0.88 &  (see GT) \\
T &  (see CA)  & 0.76/0.58 &  (see AA) & (see GA) \\
C & 3.28/2.17 & 1.80/1.45 & 1.35/1.28 &  (see GG) \\
\hline
\end{tabular}
\end{center}
\caption{
\label{tab:01}
Free energy ($\Delta G)$ of helix binding in
nearest neighbor models at 37$^o$C with
Breslauer/SantaLucia parameters (kcal/mol).
}
\end{table}

\begin{figure}[!tpb]%figure1
\begin{turn}{-90}
        \resizebox{!}{8.9cm}{\includegraphics{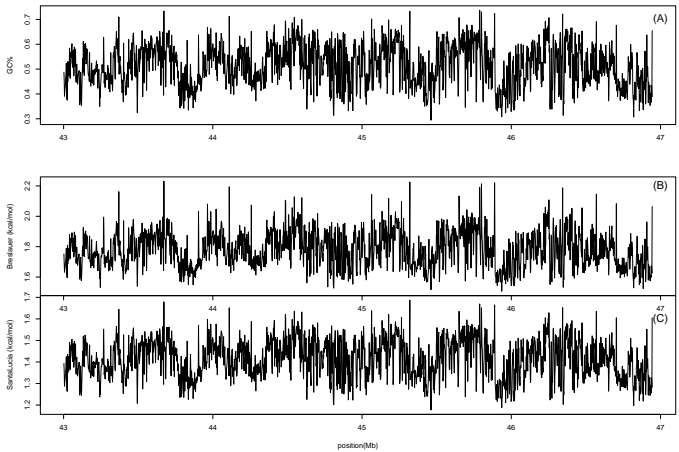}}
\end{turn}
\caption{
(A) (G+C)\% calculated in non-overlapping windows
of size 2kb;
(B) free energy $\Delta G$ in nearest neighbor model
with Breslauer's parameter values;
(C) free energy $\Delta G$ in nearest neighbor model
with SantaLucia's parameter values;
The $x$-axis is the chromosome position, in Mb.
}\label{fig:01}
\end{figure}

\SEC{DNA binding energy and stability}

It has been well known that basepairs with strong bases (G-C)
are more stable than basepairs with weak bases (A-T),
due to the presence of three versus two hydrogen bonds.
This single-base model of binding energy has been extended
to dinucleotide models where a dinucleotide step (two
neighboring basepairs) contributes an amount to the
total binding energy \citep{tinoco}. There are two commonly used
parameter value sets in the dinucleotide model: one by Breslauer
and his colleagues \citep{breslauer86} and another
summarized by SantaLucia, also known as the unified
parameters \citep{santalucia98}. The nearest-neighbor
free energy $\Delta G$ parameter values at 37$^o$C are listed
in Table \ref{tab:01} for all 16 dinucleotide steps.

A 3.9Mb sequence from the NCBI Build 35 (May'2004, hg17)
of human chromosome 21 is downloaded from the UCSC genome browser
\cite{ucsc}, starting from the position 43Mb and
ending at the right telomere, of position 46.944323Mb.

Figure \ref{fig:01} shows the (G+C)\% and averaged binding
free energy $\Delta G$ calculated by the dinucleotide model
with Breslauer's and SantaLucia's parameters, using non-overlapping
windows of 2kb. It is clear that binding energy is higher in
(G+C)\%-high peak regions, thus also oscillates with the 500kb
periodicity. However, the magnitude of oscillation is larger
in the free energy based on Breslauer's parameters than that
using SantaLucia's parameters (range of (1.51-2.23) versus
(1.18-1.69)).

Among the values of $\Delta G$ in Table \ref{tab:01}, the highest helix binding
energies are usually associated with two strong bases (G or C),
with the exception of 1.84 kcal/mol for GG/CC dinucleotide in
SantaLucia's parameters. The lowest binding energies tend to be
associated with two weak bases (A or T), but with the exceptions
of AA/TT (1.66 kcal/mol) and AT (1.19 kcal/mol)
dinucleotides in Breslauer's parameters.  The difference
between the two sets of parameters is the largest for
CG (1.11 kcal/mol, 40.7\% of the average between the two parameters),
GG/CC (0.91 kcal/mol, 39.7\%), and AA/TT (0.66 kcal/mol,
49.6\%) dinucleotides. With these exceptions,
one may not automatically assume binding energy to fluctuate
the same way as (G+C)\%. What Figure \ref{fig:01}
have shown is that the difference between the single-base
model (counting the number of weak and strong bases) and the
dinucleotide models is not large enough to destroy the 500kb
oscillation in binding energy.

The correlation coefficient between windowed energy values
and the (G+C)\% values was calculated (the first two lines in
Table \ref{tab:02}). These correlation values show that 
SantaLucia parameters are more correlated with the GC\% than Breslauer's
parameters (correlation coefficient of 0.998 versus 0.981
using the 2kb window). By examining the two sets
of free energy parameters in Table \ref{tab:01} closely, it
is clear that difference can be traced to the fact that
Breslauer's parameters assign a higher energy value for two
AT-rich dinucleotides than SantaLucia's parameters: 5'-AA-3'
and 5'-AT-3'.  It is still debatable whether Breslauer's
or SantaLucia's parameters reflect the {\sl in vivo}
situation of helix local thermodynamics \citep{mira03}, 
and the issue may not be settled soon \citep{melo05}.

\begin{figure}[!tpb]%figure2
\begin{turn}{-90}
        \resizebox{!}{8.5cm}{\includegraphics{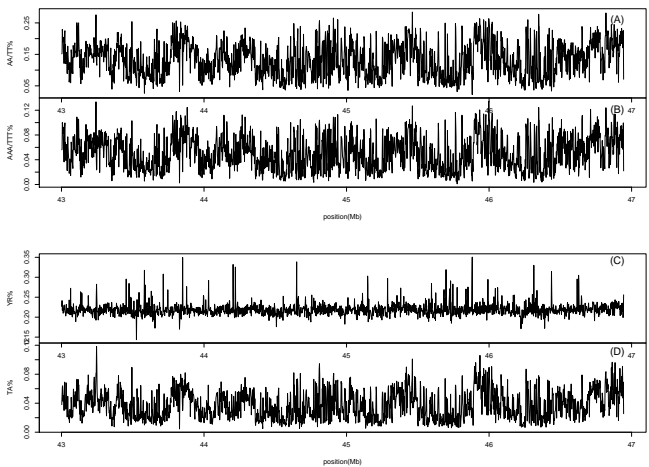}}
\end{turn}
\caption{(A): Density of AA/TT in non-overlapping windows of size 2kb;
(B) AAA/TTT density;
(C) 5'-YR-3' density;
(D) 5'-TA-3' density. }\label{fig:02}
\end{figure}

\SEC{DNA flexibility, stiffness, and untwisting}

Without an actual measurement of the DNA polymer mechanic
properties, we rely on dinucleotides and trinucleotides
that are known to be related to the DNA flexibility, stiffness, 
and untwisting to study the variation of these properties
along the chromosome.  For example, the AA..A/TT..T tract is 
known to have a stiff configuration because of an 
additional hydrogen bond between adjacent pairs along 
two diagonally located bases \citep{nelson87}. This 
hypothesis had been confirmed for AA/TT dinucleotide by 
their limited range of roll and slide values \citep{el97}. 
We use AA/TT dinucleotide and AAA/TTT trinucleotide density 
in a moving window as an indicator for the intrinsic 
stiffness of the double helix.

Unlike A/T-tracts, 5'-pyrimidine-purine-3' (5'-YR-3')
steps can adopt two possible configurations, and thus
they are flexible \citep{call04}. In a simplified
approach, we use 5'-YR-3' density as an indicator for
flexibility of the DNA double helix.

Among the four 5'-YR-3' steps (CA, CG, TA, TG),
5'-TA-3' has the weakest basepair binding.
The biconfiguration nature and weak binding make 5'-TA-3' 
one of the best candidates for untwisting initiation sites of
double helix \citep{call04}. We use the
5'-YR-3' and 5'-TA-3' density in moving windows
as an indicator for an untwisting potential.

Figure \ref{fig:02} shows densities of the above
mentioned di-/tri-nucleotide: AA/TT\%, AAA/TTT \%,
5'-YR-3' \%, and 5'-TA-3' \%.  The 500kb oscillation 
in the first two densities is clearly seen. The
5'-YR-3' density does not exhibit any regular oscillation
of 500kb, whereas 5'-TA-3'density does oscillate
with the 500kb wavelength. 

Note that the signal we are measuring by the
di-/tri-nucleotide density is different from
that of CpG island \citep{cpg}. In detecting CpG islands,
the density of 5'-CG-3' dinucleotide is normalized
by the square of GC\% (the observed over expected, or O/E),
and the presence of a signal require the 5'-CG-3'
density to be at least a quadratic function of GC\%.
In fact, it was known that the O/E signal increases
with the GC\%, indicating a cubic relationship between
5'-CG-3' density and GC\% in CpG islands \citep{matsuo93}.
Here only the ``linear" signal was measured.

\begin{figure}[!tpb]%figure3
\begin{turn}{-90}
        \resizebox{!}{8.5cm}{\includegraphics{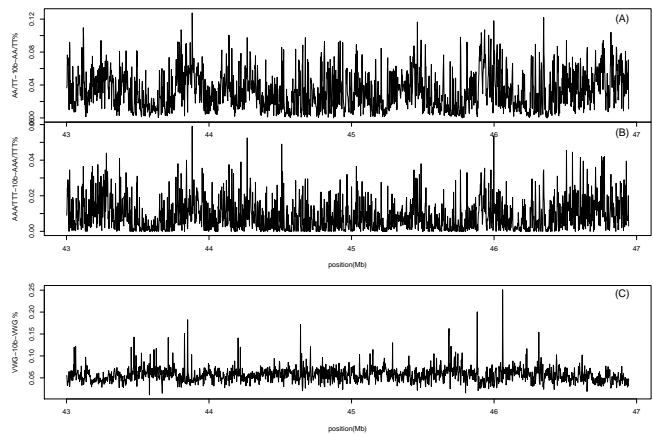}}
\end{turn}
\caption{
(A) Density of AA-10b-AA/TT-10b-TT in non-overlapping window
of size 2kb;
(B) AAA-10b-AAA/TTT-10b-TTT density.
(C) YWG-10b-VWG density, where VWG indicates [not-T][A/T][G]
or it's reverse complement triplet [C][A/T][not-A].
}\label{fig:03}
\end{figure}

\SEC{Periodicity-10-base signal and nucleosome forming 
potential}

It has been known that almost all genomes contain a AA-10b-AA/TT-10b-TT
signal \citep{period10}, where the ``10b"
can be 10 or 11 bases for individual cases, but after
averaging becomes a real number between 10 and 11.
This periodic signal is also present in the aligned
nucleosome-forming sequences \citep{sat86}.
We count the number of occurrence of AA-10-AA,
TT-10-TT, AA-11-AA, and TT-11-TT in a moving window,
then convert to density (similar calculation for
AAA-10b-AAA/TTT-10b-TTT density is also carried out).
As a crude approximation, this density is used to 
indicate the region's tendency for nucleosome
formation.

Figure \ref{fig:03} (A)(B) show the AA-10b-AA/TT-10b-TT
and AAA-10b-AAA/TTT-10b-TTT density in a 2kb non-overlapping
moving window. The 500kb oscillation is clearly seen,
and may support the idea that the nucleosome forming 
strength also oscillates with that wavelength in this region.

However, it was suggested that the regular spacing of 10 
bases of another triplet motif, [not-T][A/T][G],
can be considered as a nucleosome formation signal
(called ``VWG" signal) \citep{vwg}. We count the occurrence
of [not-T][A/T][G]-10/11-[not-T][A/T][G] and
[C][A/T][not-A]-10/11-[C][A/T][not-A] in a moving window,
whose density is plotted in Figure \ref{fig:03}(C). This
VWG signal does not exhibit a 500kb oscillation in this
region.

In a more sophisticated study based on discriminant
analysis, a composite measure called ``nucleosome
formation potential" (NFP) was proposed \cite{nfp}.
As shown in Fig.1 of \citep{vino-nfp}, this NFP value
decreases with GC\%. Since AA-10b-AA/TT-10b-TT and
AAA-10b-AAA/TTT-10b-TTT density also decreases with
GC\%, the two measures are consistent.  The VWG signal, 
however, does not have a simple relationship with 
GC\%, though mostly it increases with GC\%. 
Whether one can predict nucleosome forming potential
of a DNA sequence accurately, and whether such
an intrinsic potential really exists, seems still to be open 
questions, and it is possible that either AA/TT-10b-AA/TT
or VWG-10b-VWG signal does not present the whole
picture on nucleosome formation.

\SEC{Discussion and Conclusion}
 
Besides the helix structure related intrinsic features,
the scaffold/matrix-attached-regions (S/MARs) is another
pattern that can be determined from the DNA sequence.
S/MARs are the base/foundation of DNA loops \cite{mirk}, and
S/MAR sequences can be obtained from S/MAR databases
such as the one developed at the University of G\"{o}ttingen 
\citep{liebich}.

By examining the top 34 most frequent hexamers in S/MAR
sequences (Table 2 of \citep{liebich}(b)), it is clear
that S/MARs are AT-rich \citep{saitoh94}. In fact,
only 11 hexamers contain one G or C, ranked 10, 
16--18, 21, 22, 25--27, 29, 30 in the top34, and the rest
consist exclusively of A and T \citep{liebich}.
It is not surprising that S/MAR hexamer density
(percentage of hexamers that match the top 34
most frequent S/MAR hexamer motifs and their reverse complement)
also oscillates with a 500kb wavelenegth in this
region \citep{li-gene}.

\begin{figure}[!tpb]%figure5
\begin{turn}{-90}
        \resizebox{8cm}{8cm}{\includegraphics{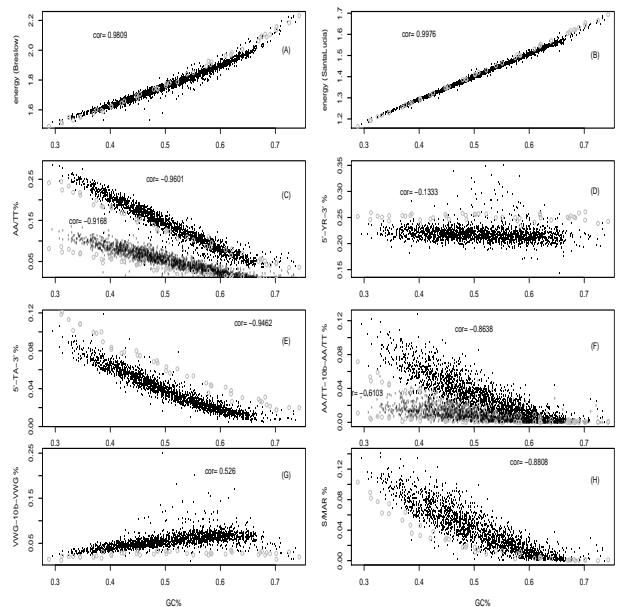}}
\end{turn}
\caption{
Scatter plots of ten quantities versus GC\%: 
(A) helix binding energy by Breslauer's model;
(B)  binding energy by SantaLucia's model;
(C) AA/TT (upper) and AAA/TTT (lower, using the symbol ') densities; 
(D) 5'-YR-3' density; 
(E) 5'-TA-3' densities;
(F) AA-10b-AA/TT-10b-TT (upper) and AAA-10b-AAA/TTT-10b-TTT 
(lower, using the symbol ') densities;
(G) VWG-10-VWG densities;
and (H) density of the top 34 hexamers in
known S/MAR sequences and their reverse complements.
The corresponding values for randomized sequences
are also shown (grey circles). The correlation coefficient
between these quantities andf GC\% is indicated on the plot.
}\label{fig:05}
\end{figure}

The existence of 500kb oscillation in most of the 
quantities we have examined indicates that these
structure-related sequence features are correlated with GC\%.
To assess this correlation directly, Figure \ref{fig:05} shows the
scatter plot of ten quantities used in Figures 1-3 as versus
GC\%,  and Table \ref{tab:02} lists correlation coefficients of
all pairs among these eleven quantities. Figure \ref{fig:05} and
Table \ref{tab:02} have confirmed that these structure-based
sequence features are highly correlated (test results of these
correlation coefficients are all significant with the exception
of a few pairs involving 5'-YR-3'), and GC\% can be 
used as a good surrogate for these features (with
the exception of 5'YR-3').

Density of 5'-YR-'3 is not correlated with other quantitied
studied (4 correlation coefficients are not significant at
the $p$-value=0.01 level, and 5 other correlation coefficients,
though significant, are rather weak). The next group of
quantities that have weak correlation with others are
the AAA-10b-AAA/TTT-10b-TTT and VWG-10b-VWG densities,
with several correlation coefficients in the 0.4-0.5 range.

\begin{table*}[!t]
\begin{center}
\begin{tabular}{lcccccccccc}
 & GC & Breslauer & SantaLucia & 5'YR3' & AA & AAA & 5'TA3' & AA10AA & AAA10AAA  & VWG10VWG   \\
\hline
Breslauer & 0.981&        &       &  & & & & & &  \\
SantaLucia& 0.998&  0.985&        & & & & & & &  \\
5'YR3'   & -0.133& -0.195& -0.103&  & & & & & &  \\
AA       & -0.960& -0.896& -0.950& -0.042* &  & & & & &  \\
AAA      & -0.917& -0.844& -0.903& -0.044* & 0.974& &  & & &  \\
5'TA3'   & -0.946& -0.915& -0.947&  0.183  & 0.912& 0.858&  & & &  \\
AA10AA   & -0.864& -0.791& -0.851& -0.043* & 0.922& 0.956& 0.810&  & &  \\
AAA10AAA & -0.610& -0.545& -0.595& -0.064  & 0.683& 0.789& 0.557& 0.866&  & \\
VWG10VWG & 0.526 & 0.398 & 0.514 & 0.279 & -0.657 & -0.637 & -0.574 & -0.601 & -0.458  &\\
S/MAR    & -0.881& -0.807& -0.868& -0.002* & 0.929& 0.967& 0.854& 0.947& 0.810 & -0.617\\
\hline
\end{tabular}
\end{center}
\caption{
\label{tab:02}
Correlation coefficients of eleven quantities obtained from non-overlapping
2kb windows: GC\%, bindinger energy by Breslauer's model and SantaLucia's
model, densities of 5'-YR-3', AA/TT, AAA/TTT,
5'-TA-3', AA-10b-AA/TT-10b-TT, AAA-10b-AAA/TTT-10b-TTT,
VWG-10b-VWG, and density of top S/MAR hexamers.
Testing of correlation coefficient equal to zero is significant
at $p$-value=0.01 level for all pairs except those marked by
the stars (YR-AA $p$=0.064, YR-AAA $p$=0.049, YR-AA10AA $p$=0.056, 
and YR-SMAR $p$=0.93).
}
\end{table*}

One may ask the question on whether the correlation between
these quantities and GC\% is ``trivial", because these
patterns are either dominated by GC-rich or AT-rich 
di- tri-nucleotides. This question can be addressed by
examining the GC\%-preserving random sequences. In
Figure \ref{fig:05} the ten structure-related quantities
for the random sequences are shown as a function of GC\%
(circles). Several interesting observations can be made.

\begin{itemize}
\item
Binding energies calculated on real DNA sequences are very
close to those calculated on randomized sequences. However, 
the binding energy of real DNA sequences is slightly
lower than that of random sequences at high GC\% values.
A similar observation was made in \citep{vino} (Fig.1(C)
of  \citep{vino}) on the ``relative" thermostability.
\item
The A/T-tract density is higher in real DNA sequences
than randomized sequences, mainly in the AT-rich ranges.
It indicates that DNA sequences are more rigid than
randomized sequences in general.
\item
The biconfigurational 5'-YR-3' dinucleotide density is lower
in real DNA sequences than randomized sequences
(with some exceptions for DNA segments with GC\%
around 50\%-60\%). It indicates DNA sequences are
less flexible than randomized sequences.
\item
The 5'-TA-3' density is lower in DNA sequences than
random sequences, making them less susceptible
for helix untwistings.
\item
The periodicity of 10/11 bp signal for both AA/TT, AAA/TTT,
and VWG triplet has a stronger presence in real DNA
sequences than random sequences, probably making them more
likely to form nucleosomes.
\item
The S/MAR potential is higher in DNA sequences than
randomized sequences.
\end{itemize}

From these observations, one may expect that the
binding energy faithfully follows the same variation 
and oscillation as GC\%; A/T tract density, TA density,
AAA-10b-AAA signal, and S/MAR signal more or less
follow the same oscillation as GC\%; YR density,
AAA-10b-AAA signal, and YWG-10b-YWG signal may not
follow the same oscillation as GC\%.

It has been known that GC\% conveys biological information \citep{isochore}(c).
For example, the Giemsa-dark chromosome staining band, or G-band,
is AT-rich, whereas the Giemsa-light band or R-band is GC-rich
\citep{ikemura88}, or by a new hypothesis, AT-rich and
GC-rich relative to its neighboring bands \citep{gojobori02}.
Gene density is another example, with GC-rich regions being
relatively gene-rich  \citep{mouch91}.  Fluorescence microscopy
images show that chromosomes inside the nucleus are organized
in a radial order, called ``chromosome territories"
\citep{cremer}. The GC-rich, gene-rich regions
tend to be located towards the center of the nucleus \citep{saccone02}, and
the corresponding chromatin compartments are more
``open" \citep{cremer}.

Without experimental evidences, it is difficult to speculate what
type of high-order chromatin structure this 500kb oscillation
might cause. According to the chromatin structure model
summarized in \citep{filipski90}, there could be multiple
level of foldings in the hierarchical structure of a chromatid:
Watson-Crick's double helix (10bp for one helix turn),
nucleosomes ($\sim$ 200bp per unit), solenoids (6
nucleosome units per helix turn, or 1.2kb) that twist
to form a loop of $\sim$ 50kb, rosettes that consist
of 6 loops ($\sim$ 300 kb), coils that consist of 30
rosettes ($\sim$ 9Mb), and finally the chromatids
consist of, for a medium sized human chromosome,
$\sim$ 10 coils. Within the framework of
this model, our 500kb oscillation matches roughly the size
of a rosette. However, we should caution that the
exact figure for the size of these hierarchical units
is illustrative, and the model itself may be too much
based on {\sl in vitro} experiments, and on inactive
cells \citep{vanholde95}.

The unique large-scale oscillation of GC\% in human
chromosome 21 studied in this paper and in \citep{li-holste04} 
can be further analyzed from several perspectives.
One is about its evolutionary presevation in other
species. Due to the high degree of similarity between
human and chimpanzee, it is natural to assume that the
same 500kb oscillation would also be present in chimpanzee
genome. Indeed, it was shown that 500kb oscillation
exists in chimpanzee chromosome 22 \cite{li-gene}. On
the other hand, no such 500kb oscillation was observed
in mouse genome. It would be interesting to check its
existence in species in-between mouse and human.

It was suggested for the yeast genome \citep{filipski02}
that the transcription direction of open reading frame (ORF)
points from GC-rich to GC-poor regions. Combined with
the general picture that DNA loop anchored in AT-rich
regions whereas the GC-rich part of the loop is exposed 
to the outside, transcription likely starts from the 
top of DNA loop to loop base. Although the length scale 
between two GC-rich regions analyzed in the yeast genome 
($\sim$10kb) is much shorter than the GC\% oscillation length
studied here, there are some evidence
that gene density on two opposite strands alternating
in this region (Fig.5(c) of \cite{li-holste04}). A more
careful analysis is needed to confirm the similarity
between human and yeast genome, and the regular oscillation
of GC\% discussed here provides an ideal test ground.

In conclusion, the 500kb oscillation in GC\% as reported in 
\citep{li-holste04} was shown to lead to similar oscillation
of some intrinsic structure-related patterns. And we
hypothesis that a regular oscillation in chromatin structure 
with the same wavelength is also present in this
region.

\SEC*{Acknowledgements}

W.L. acknowledges the financial support at the
The Robert S Boas Center for Genomics and Human Genetics.
P.M. thanks the support of DGAPA project IN111003.

%%%%%%%%%%%%%%%%%%%%%%%%%%%%%%%%%%%%%%%%%%%%%%%%%% bibliography \bibitem [] {}

\end{document}